\documentclass{aa}
\usepackage{natbib} 
\usepackage{graphicx}
\usepackage{txfonts}
\usepackage{epstopdf}

\begin{document}

\title{Multi-epoch VLBA observations of radio galaxy 0932+075:\\
       is this a compact symmetric object?}
\titlerunning{Is radio galaxy 0932+075 a compact symmetric object?}

\author{Andrzej Marecki\inst{1} \and Aleksandra Soko{\l}owska\inst{1,2}}

\offprints{Andrzej Marecki}

\institute{Toru\'n Centre for Astronomy, Faculty of Physics, Astronomy and
           Informatics, Nicolaus Copernicus University, Toru\'n, Poland\\
           \email{amr@astro.uni.torun.pl}
      \and
           Institute for Computational Science, University of Zurich,
           Winterthurerstrasse 190, 8057 Z\"urich, Switzerland\\
           \email{alexs@physik.uzh.ch}
           }

\date{Received 15 July 2013 / Accepted 4 July 2014}

\abstract{A part of the radio structure of the galaxy 0932+075 emerged as a 
possible compact symmetric object (CSO) after the observation with the Very 
Long Baseline Array (VLBA) at 5\,GHz in 1997. More than a decade later, we 
carried out observations at 5, 15.4, and 22.2\,GHz using the VLBA to test 
this possibility. We report here that we have found a component whose 
spectrum is inverted in the whole range from 5\,GHz to 22\,GHz and we label 
it a high-frequency peaker (HFP). Using a set of 5\,GHz images from two 
epochs separated by 11.8\,years and a set of 15.4\,GHz images separated by 
8.2\,years, we were able to examine the proper motions of the three 
components of the CSO candidate with respect to the HFP. We found that their 
displacements cannot be reconciled with the CSO paradigm. This has led to 
the rejection of the hypothesis that the western part of the arcsecond-scale 
radio structure of 0932+075 is a~CSO anchored at the HFP. Consequently, the 
HFP cannot be labelled a core and its role in this system is unclear.
}


\maketitle

%________________________________________________________________

\section{Introduction}

The term compact symmetric object (CSO) was introduced by
\citet{Readhead1994} and \citet{Wilkinson1994} to label three objects 
from the survey of 65 very compact radio sources observed using Very Long 
Baseline Interferometry (VLBI) at 5\,GHz \citep{PR1988}. The authors pointed 
out that despite their small sizes ($\lesssim$1\,kpc) CSOs were symmetric 
and in this regard they resembled much larger radio sources. The nature of 
CSOs could be explained in two ways that were immediately taken into 
consideration by \citet{Readhead1994}: CSOs are either precursors of 
standard large-scale doubles or they constitute a class of short-lived 
objects that decay too quickly to make it possible for them to evolve from 
subkiloparsec-sized structures to more extended forms. Both options belong 
to the so-called youth scenario of compact sources. It appears that since 
1994, when the CSO class was recognised, the youth scenario has been valid 
\citep[see e.g.][for a review]{Fanti2009}.

The most straightforward proof that CSOs are young is based on the 
measurements of kinematic ages calculated from their lobe expansion
velocities extracted from multi-epoch VLBI observations. These measurements 
were carried out by \citet{OC98}, \citet{Owsianik1998}, \citet{Taylor2000}, 
\citet{Marecki2003}, \citet{PC2003}, \citet{Gugliucci2005},
\citet{Nagai2006}, \citet{Polatidis2009}, and \citet{An2012}. To date, the 
lobe expansion velocities, or at least their lower limits, have been found 
for 37\,CSOs and so their kinematic ages, or at least their upper limits, 
could have been estimated. Given that for 27 of them the redshifts are 
known, both lobe expansion velocities and the kinematic ages of the sources 
can be calculated properly, i.e. with the time-dilation factor taken into 
account. They happen to be less than 3000\,years old, but a considerable 
fraction -- eleven CSOs out of the twenty-seven -- are younger than 
500\,years, which is a meaningful overabundance pointed out by 
\citet{Gugliucci2005}. This circumstance supports the conjecture that 
the evolution of the classic double radio sources often comes 
to a premature end at the CSO stage as suggested by \citet{Readhead1994}.

\begin{figure*}
\centering 
\resizebox{\hsize/2-0.85cm}{!}{\includegraphics{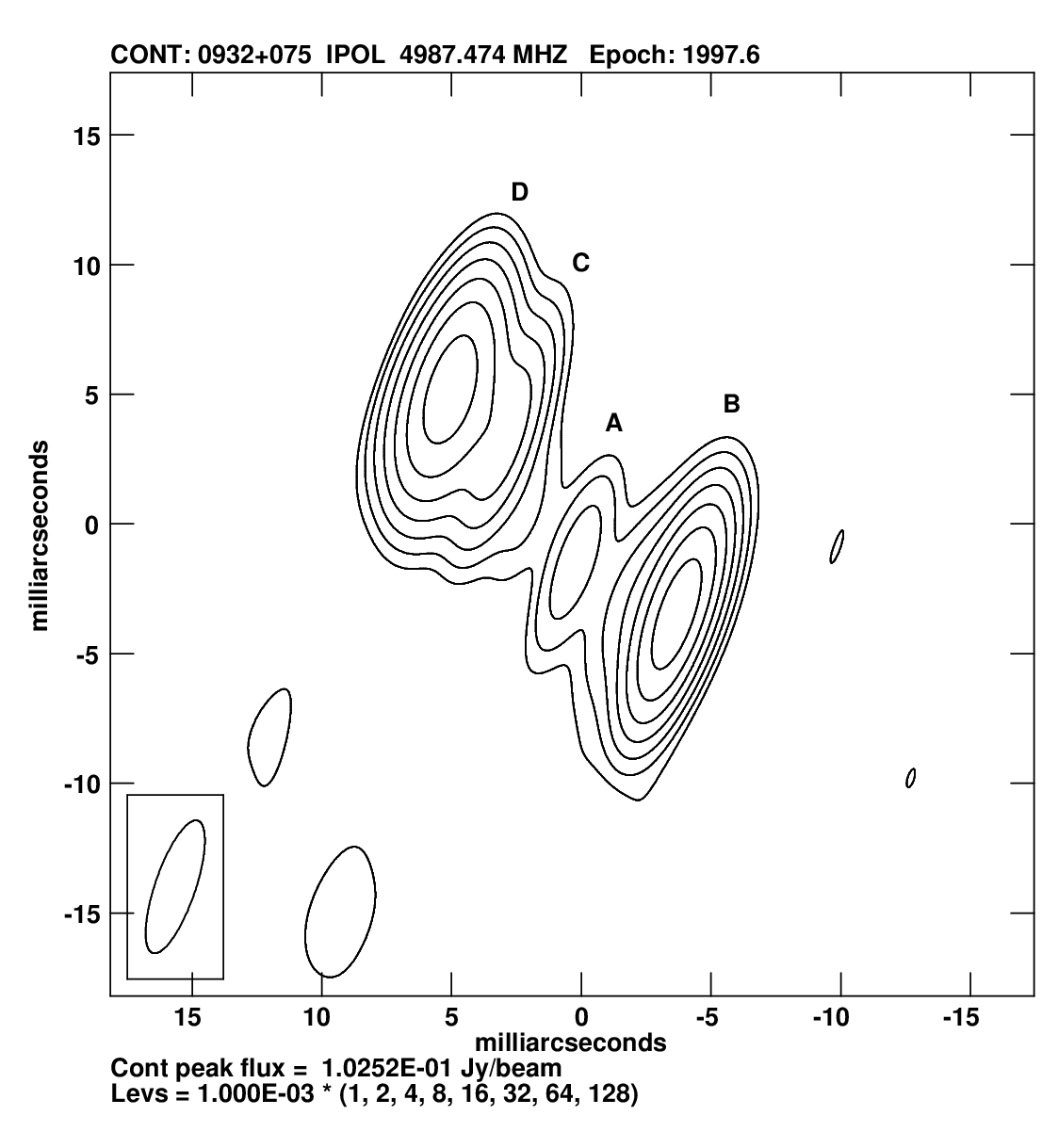}} 
\resizebox{\hsize/2-0.85cm}{!}{\includegraphics{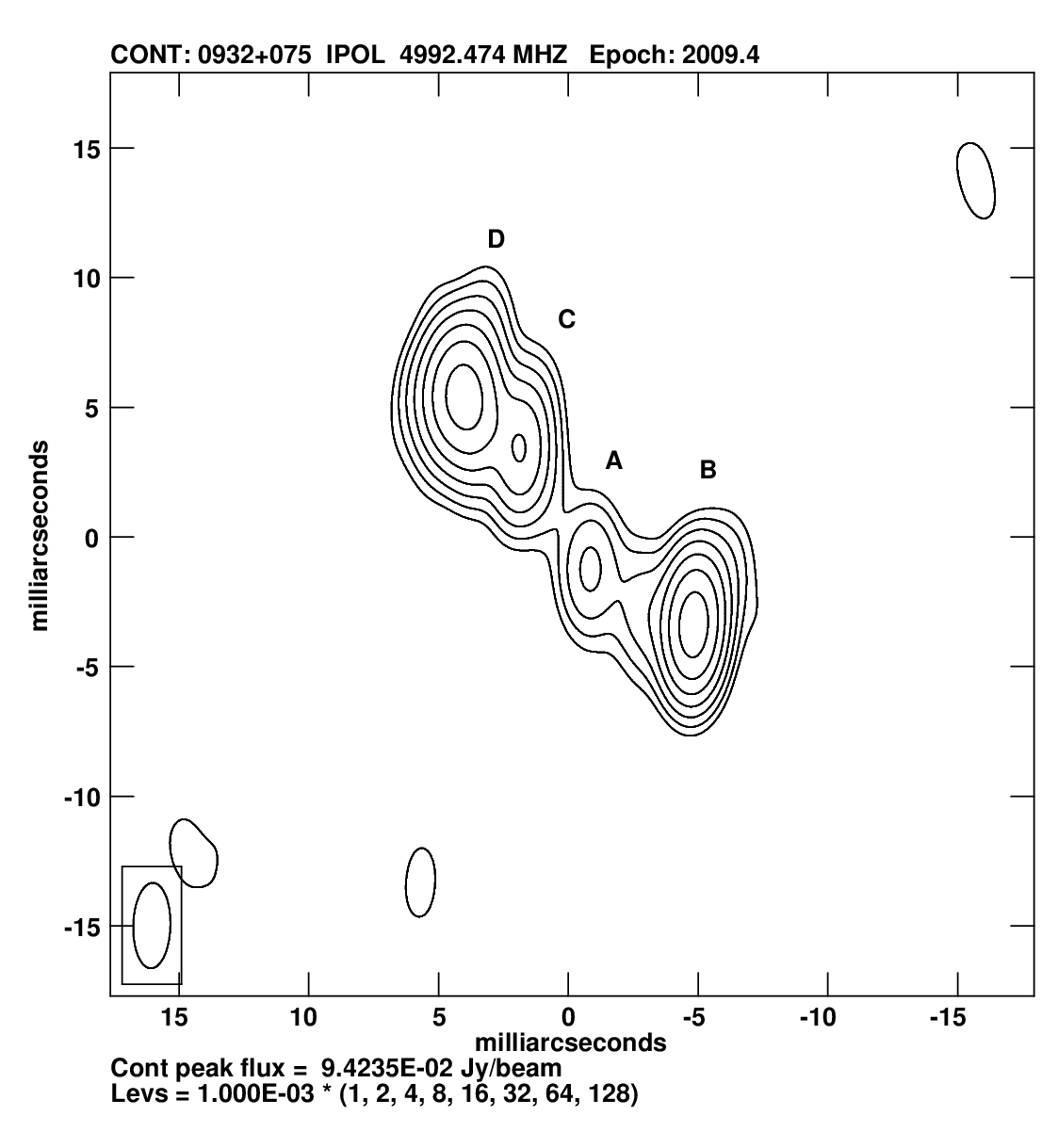}} 
\resizebox{\hsize/2-0.85cm}{!}{\includegraphics{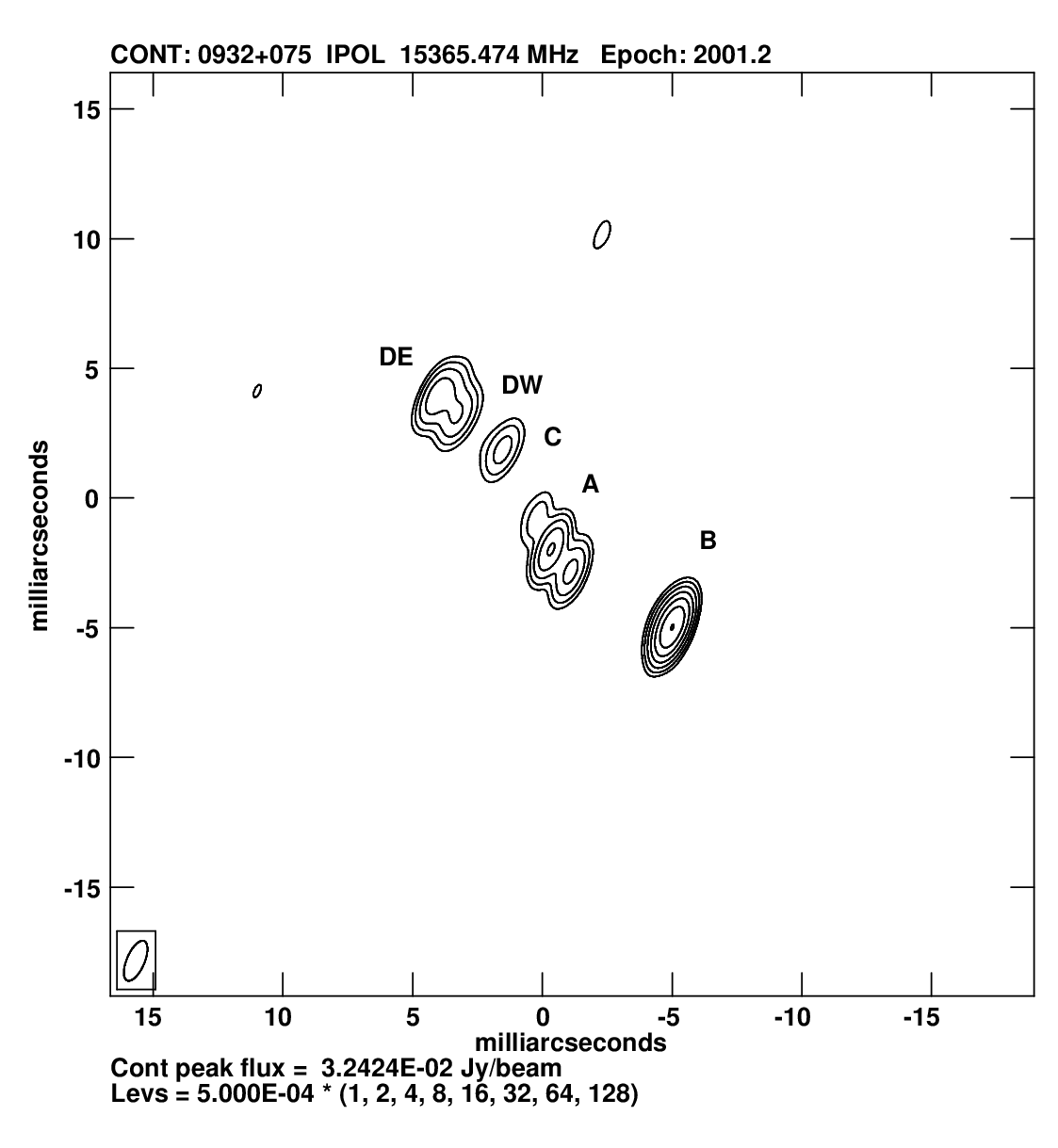}} 
\resizebox{\hsize/2-0.85cm}{!}{\includegraphics{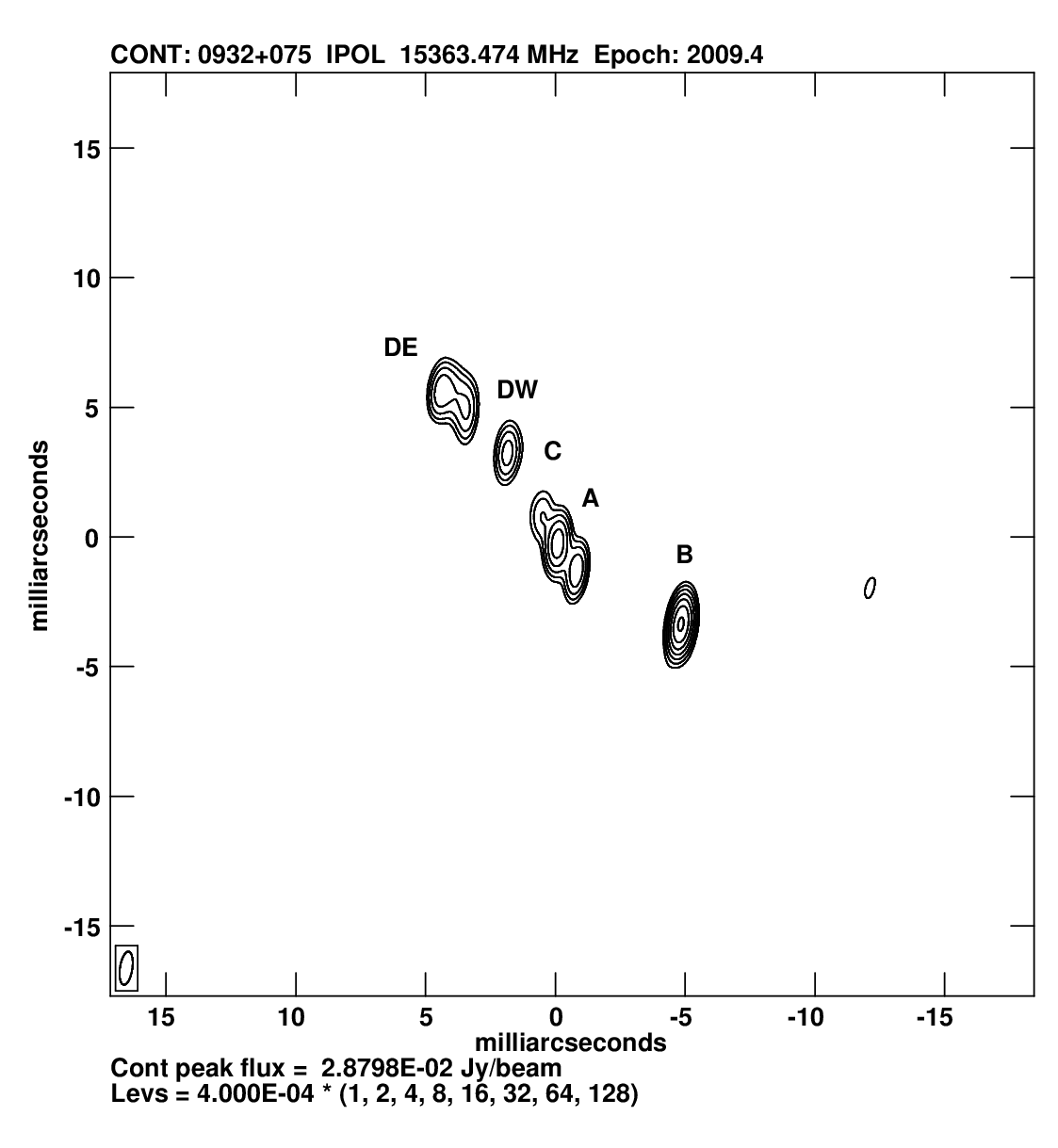}}
\caption{The western part of 0932+075 as seen with the VLBA. In all panels, 
the contours are increased by a factor of 2. {\em Upper left:} The image 
resulting from the 5\,GHz observation carried out on 02~August~1997 and 
reported in \citet{Browne2003}. The first contour level corresponds to 
1\,mJy/beam. The beam size is $5.4 \times 1.6$\,mas at the position angle of 
$-19\degr$. {\em Upper right:} The image resulting from the 5\,GHz observation
that we carried out on 17~May~2009. The first contour level corresponds to
1\,mJy/beam. The beam size is $3.3 \times 1.4$\,mas at the position angle of 
$-1.6\degr$. {\em Lower left:} The image resulting from the observation 
carried out at 15.4\,GHz on 22~March~2001. The first contour level
corresponds to 0.4\,mJy/beam. The beam size is $1.7 \times 0.7$\,mas at the
position angle $-23\degr$. {\em Lower right:} The image resulting from the 
15.4\,GHz observation that we carried out on 17~May~2009. The first contour
level corresponds to 0.4\,mJy/beam. The beam size is $1.3 \times 0.5$\,mas at
the position angle of $-8.5\degr$.
}
\label{fig:VLBA_obs_5_15}
\end{figure*}

Since CSOs are an astrophysically important albeit not very numerous class 
of objects, the original idea of the investigation presented here was to test 
whether \object{0932+075} (JVAS\,J0935+0719), a radio source identified with 
a galaxy at $z=0.29$ that has not yet been recognised as a CSO in the
literature, could in fact be another CSO. The scientific rationale of such an
endeavour was that 0932+075 was mapped with the VLA at 8.4\,GHz in 
the course of the Jodrell Bank-VLA Astrometric Survey (JVAS) 
\citep{Browne1998} and then included in JVAS although with a caveat that it 
was not a point-like source but an asymmetric double whose components 
are separated by $\sim$$0\farcs4$. It was therefore selected as
a~gravitational lens (GL) candidate and followed up with MERLIN at 5\,GHz in 
the course of the Cosmic Lens All-Sky Survey (CLASS) \citep{Myers2003}. 
These observations confirmed that 0932+075 was a $0\farcs38$-wide, i.e. 
1.6\,kpc-wide\footnote{For redshift $z=0.29$ and the standard cosmological 
parameters ($H_0=71~{\rm km~s}^{-1} {\rm Mpc}^{-1}, \Omega_{\rm m}=0.27, 
\Omega_{\Lambda}=0.73$), the angular distances pertinent to 0932+075 should 
be multiplied by a factor of 4.317\,pc/mas to be converted to projected 
linear distances.} double. As such, it was still a GL candidate and 
was re-observed at 5\,GHz with the VLBA on 02~August~1997 (epoch 1997.6).
Based on the results of that observation, the possibility of the presence of 
a~GL system appeared unlikely because the surface brightness values of the 
two major components were highly unequal: their ratio was 22.6:1 
\citep{Browne2003}. However, a new circumstance emerged at this point and 
subsequently became the cornerstone of the present work. Owing to the 
resolution attainable with the VLBA, the detailed structure of 0932+075 was 
revealed \citep[see Fig.\,8 in][]{Browne2003}. In particular, the western 
feature shown in that image, i.e. the component that is $\sim$23 times 
stronger than its eastern companion, turned out to be an almost symmetric 
compact double. Given its overall span ($\sim$50\,pc), it was justified to 
tentatively label the western part of 0932+075 a~CSO. Interestingly, the 
position angle of the potential CSO (45$\degr$) is not the same as that of 
the line connecting the two arcsecond-scale components (60$\degr$).

The raw observational data acquired by \citet{Browne2003} belongs to the 
public domain. We took this opportunity and carried out a standard 
reduction procedure of that data in AIPS independently. The structure of 
the CSO-like part that we have obtained is shown in the upper-left panel of 
Fig.\,\ref{fig:VLBA_obs_5_15}. In that image, three conspicuous components: 
A, B, and D and at least one more, C, are visible, although C is blended with 
its stronger neighbouring component D. The interpretation of the three 
well-resolved components is not obvious, however. While the dominant 
features at the source's extremities are likely to be the lobes, the nature 
of the third one between them is unclear; it could be the core, but 
this is only a hint that cannot be regarded as a piece of evidence in 
favour of such a conjecture. Nevertheless, since the suggestion that the 
structure shown in the upper-left panel of Fig.\,\ref{fig:VLBA_obs_5_15} 
could be a CSO seemed reasonable, we launched an observing programme to 
either prove it or reject it.

\begin{figure}
\centering 
\resizebox{\hsize-0.85cm}{!}{\includegraphics{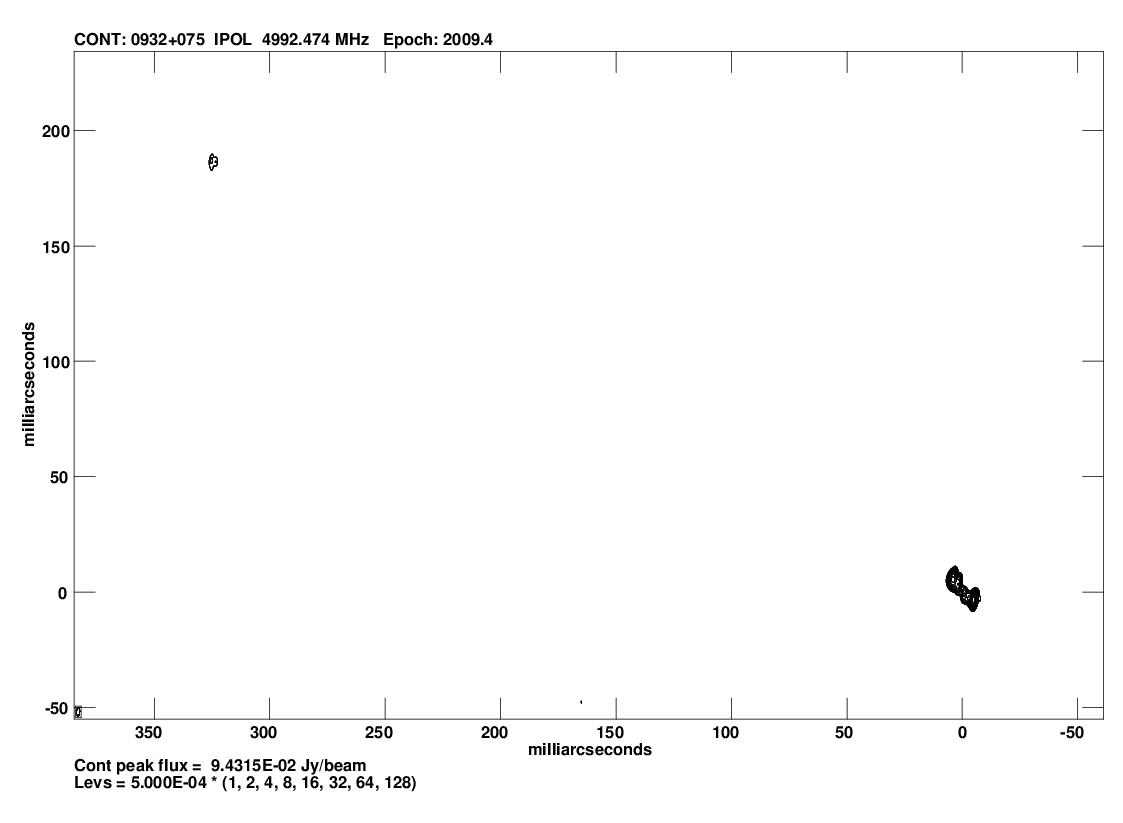}} 
\caption{The complete image of 0932+075 at 5\,GHz resulting from the VLBA
observation we carried out on 17~May~2009. The first contour level 
corresponds to 0.5\,mJy/beam. Contours are increased by a factor of 2.}
\label{fig:VLBA_obs_5}
\end{figure}

\section{New high-resolution observations}
\label{sect:epoch2}

We carried out the second-epoch observations of 0932+075 with the VLBA 
at two frequencies, 5\,GHz and 15.4\,GHz, on 17~May~2009 (epoch 2009.4), 
i.e. 11\,years and 9.5\,months after the first epoch. Sixty-four-MHz 
bandwidth was sampled at both frequencies (LHC polarisation). The first 
frequency was chosen for the sake of compatibility with the data used by 
\citet{Browne2003} so that possible proper motions of the components could 
be found, whereas the observation at the second was requested to estimate
spectral indices aiming at identification of the putative core. The source
0932+075 was observed for 80\,min at each frequency. No phase referencing 
was applied\footnote{Browne et al. did not use phase referencing either.} 
as the source was strong enough to obtain the fringe residual delays 
and rates using its own visibilities. Standard continuum processing was 
applied. The resulting images are presented in
Figs.\,\ref{fig:VLBA_obs_5_15} and\,\ref{fig:VLBA_obs_5}. The 5\,GHz image in 
Fig.\,\ref{fig:VLBA_obs_5} is a complete view of the source. It encompasses 
both its parts: the weak and diffuse eastern component and the dominant 
western one. Figure\,\ref{fig:VLBA_obs_5} is an equivalent of Fig.\,8 in 
\citet{Browne2003}; it shows exactly the same part of the sky at the same 
frequency, but in an 11.8-year later epoch. The western part of 0932+075 
alone, as seen in epoch 2009.4, is shown in the upper-right (5\,GHz) and 
lower-right (15.4\,GHz) panels of Fig.\,\ref{fig:VLBA_obs_5_15}.

\begin{figure}
\centering
\resizebox{\hsize-0.85cm}{!}{\includegraphics{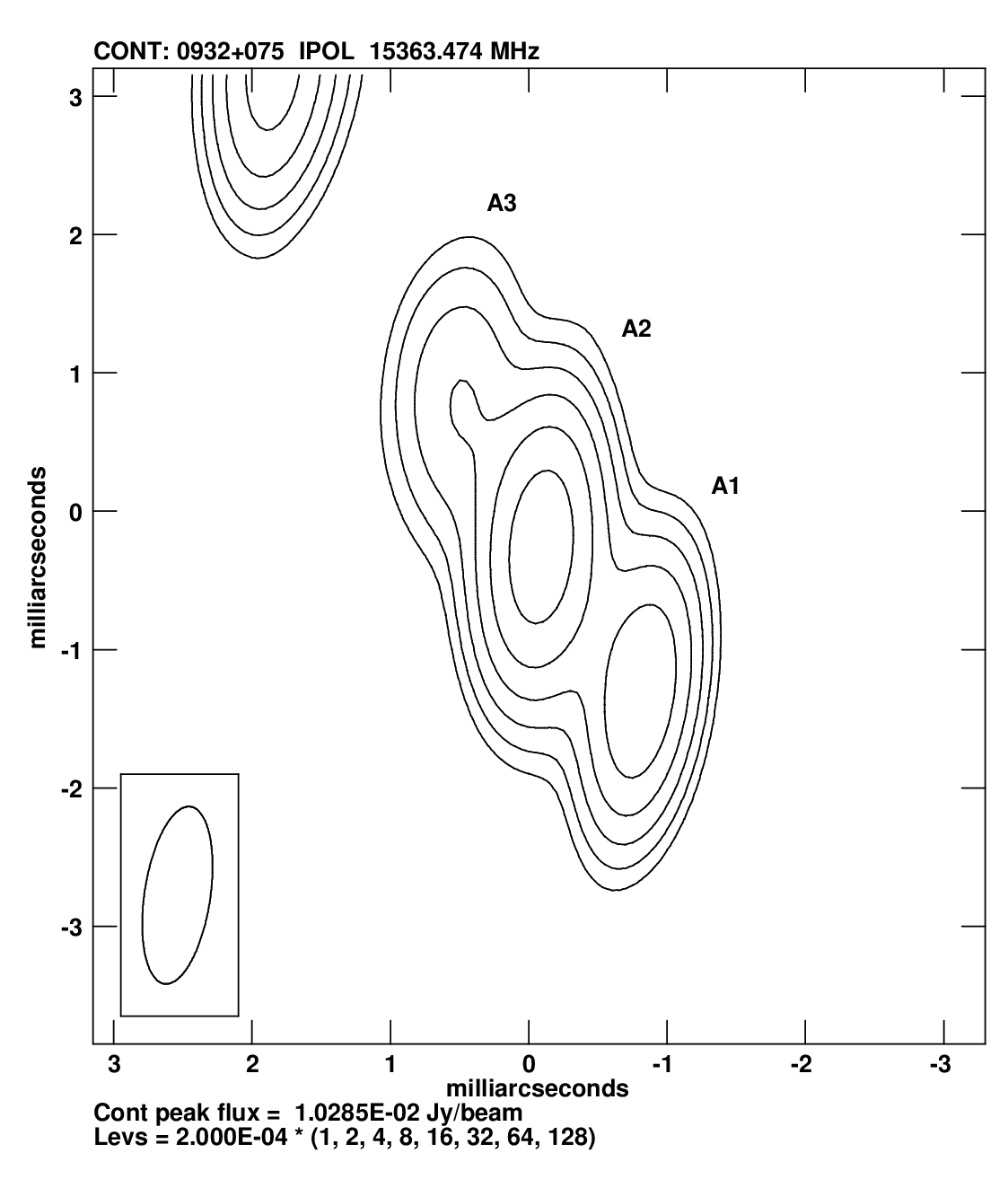}}
\caption{Enlarged cutout from the 15.4\,GHz image shown in 
Fig.\,\ref{fig:VLBA_obs_5_15} (lower-right panel) centred on the A-complex.}
\label{fig:core}
\end{figure}

In the epoch 2009.4 5\,GHz image, the western part of 0932+075 consists
of four well-resolved components. The outer two are dominating and quite 
symmetric -- their flux densities amount to 100 and 132\,mJy (see 
Table\,\ref{tab:fluxes}) -- hence the CSO nature of the western part of 
0932+075 seems to be plausible at first sight. The interpretation of the 
inner two components in that image is not straightforward though: either 
one of them is the core or neither is. The 5\,GHz and 15.4\,GHz flux densities 
of C in epoch 2009.4 are 44\,mJy and 5.4\,mJy, respectively (see 
Table\,\ref{tab:fluxes}), which means that C can be excluded as a 
possible core due to its very steep spectrum. Next, we investigate whether 
the core is associated with the component A. However, while unresolved at 
5\,GHz, this component reveals a triple structure at 15.4\,GHz, so at
that stage the location of the core remains an open matter. 

Figure\,\ref{fig:core} presents a magnified view of the A-complex extracted 
from our 15.4\,GHz image in Fig.\,\ref{fig:VLBA_obs_5_15} (lower-right 
panel). We overlaid the 5\,GHz and 15.4\,GHz images of the epoch 2009.4 to 
identify the respective components using B as a reference point because of 
its compactness. As a result, the positions of C at both frequencies matched 
well, the position of D at 5\,GHz was halfway between DE and DW at 
15.4\,GHz, whereas the component A at 5\,GHz matched neither A2 nor A3, but 
its position agreed with that of A1 at 15.4\,GHz.

\begin{figure}
\centering
\resizebox{\hsize-0.85cm}{!}{\includegraphics{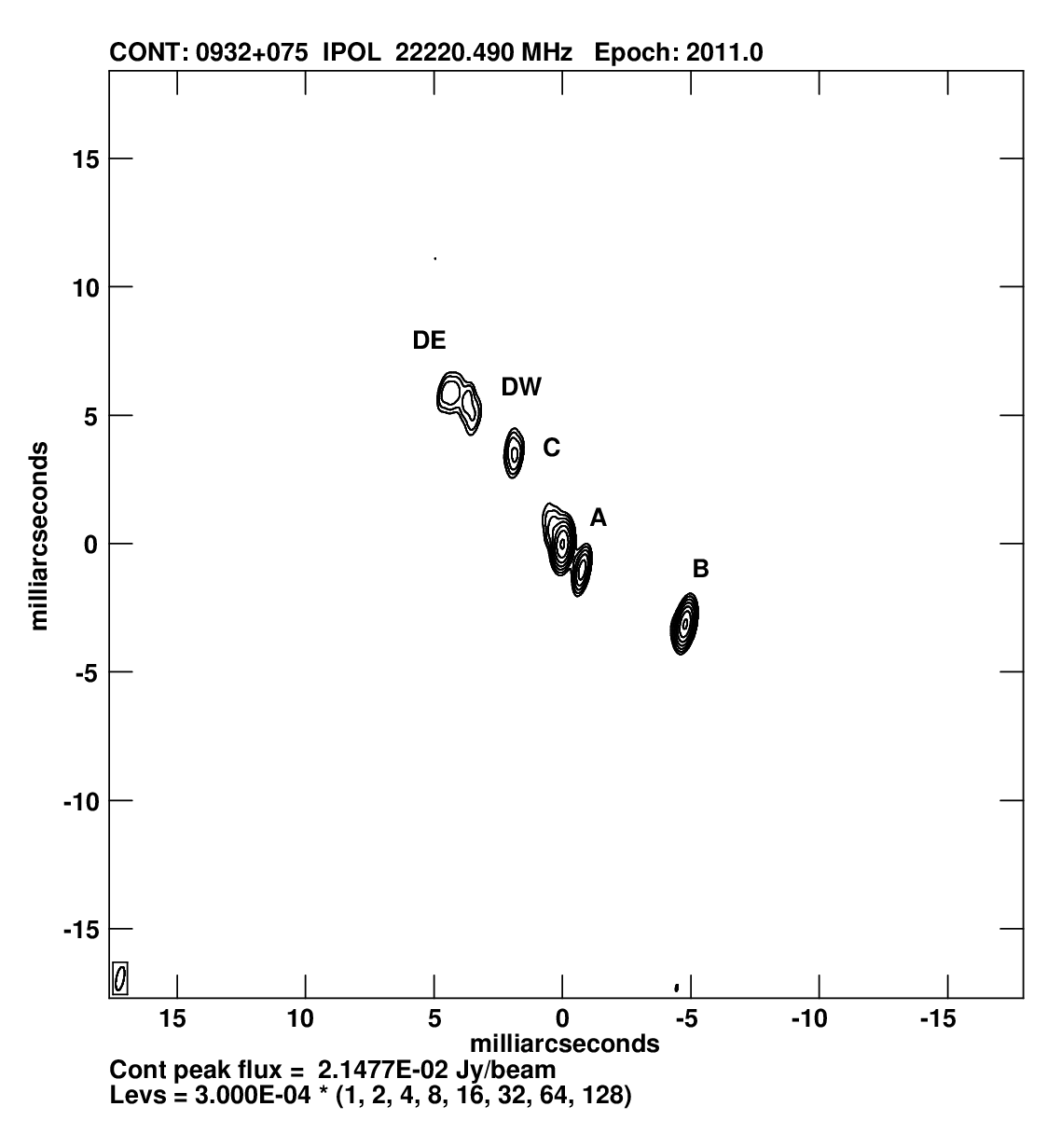}} 
\caption{The image resulting from the observation carried out at 22.2\,GHz 
on 15~December~2010. The first contour level corresponds to 0.3\,mJy/beam. 
The beam size is $0.9 \times 0.3$\,mas at the position angle of $-11\degr$.}
\label{fig:VLBA_obs_22}
\end{figure}

To confirm this finding quantitatively, we fitted Gaussian profiles to A and 
B at 5\,GHz as well as to A1, A2, A3, and B at 15.4\,GHz using AIPS task 
JMFIT. We found that the separations B--A at 5\,GHz and B--A1 at 15.4\,GHz 
were the same (4.5\,mas) while the separations B--A2 and B--A3 were 5.7\,mas 
and 6.6\,mas, respectively. We assume that the uncertainties of these 
figures amount to 20\% of the beam size\footnote{This is much more than the 
formal position errors given by JMFIT; JMFIT, however, does not take into 
account either the density of the $u-v$ plane coverage, or the errors of the 
individual visibilities. The actual errors are very difficult to estimate 
hence our very conservative assumption of their magnitude.}. Clearly then, 
component A seen at 5\,GHz is not a counterpart of either A2 or A3 seen at 
15.4\,GHz. This immediately raises the question whether any counterpart of 
A2 is present in the 5\,GHz map. It appears that there is no trace of such a 
counterpart in the 5\,GHz image, although in principle it could exist 
because, given the separation of 1.2\,mas between A2 and A1 as measured at 
15.4\,GHz, the resolution attained at 5\,GHz is sufficient to resolve these 
two components at least partly. It follows that the flux density of A2 is 
too low at 5\,GHz to make it visible hence its spectrum between 5\,GHz and 
15.4\,GHz is surely inverted. It could thus be expected that it might also 
be inverted towards higher frequencies.

To test this conjecture, we observed the western part of 0932+075 with the 
VLBA at 22.2\,GHz. The observation was carried out on 15~December~2010 
(epoch 2011.0) covering a 128 MHz bandwidth (LHC polarisation). The 
target was observed for 220\,min, again without phase referencing as the 
source was strong enough to obtain the fringe residual delays and rates 
using its own visibilities. Standard continuum processing was applied. The 
resulting image is shown in Fig.\,\ref{fig:VLBA_obs_22}.

Finally, we searched the NRAO archive to find whether there existed any VLBA 
observations of 0932+075 other than those by \citet{Browne2003} and ours. We 
found that 0932+075 had been observed at 15.4\,GHz on 22~March~2001 (epoch 
2001.2), but the outcome of that observation was never published (see {\it 
Acknowledgements} for details). The total on-source time was $\sim$70\,min 
for 0932+075 at 32 MHz bandwidth with LHC polarisation only. We processed 
the public-domain raw data acquired in the course of that observation in a 
standard way. The resulting image is shown in Fig.\,\ref{fig:VLBA_obs_5_15} 
(lower-left panel).

\begin{table}
\caption{Flux densities of the components of the western part of 0932+075 [mJy]}
\label{tab:fluxes}
\centering
\begin{tabular}{l r r r r r}
\hline
\hline
Epoch  & 1997.6 & 2009.4 & 2001.2 & 2009.4 & 2011.0\\
\hline
Component & \multicolumn{5}{c}{Frequency [GHz]}\\
 & 5.0 & 5.0 & 15.4 & 15.4 & 22.2\\
\hline
A & 7.1 &  11  &  -- &  -- & --\\
A1 & -- &  --  & 6.8 & 6.6 & 8.3\\
A2 & -- &  --  & 9.3 &  12 & 31\\
A3 & -- &  --  & 4.0 & 3.7 & --\tablefootmark{a}\\
B & 108 & 100 & 36  & 32  & 28\\
C & 37  & 44  & 3.9 & 5.4 & 5.2\\
D & 141 & 132 & --  & --  & --\\
DW & -- & --  & --\tablefootmark{b} & 8.2 & 6.4\\
DE & -- & --  & 11  & 8.4 & 5.1\\
\hline
\end{tabular}
\tablefoot{\\
\tablefoottext{a}{Flux density of the component A3 could not be measured reliably
because of blending with neighbouring component A2.}\\
\tablefoottext{b}{Flux density of the component DW could not be measured reliably
because of blending with neighbouring component DE.}\\
}
\end{table}

We measured the flux densities of the components of 0932+075 in all five 
maps shown in Figs.\,\ref{fig:VLBA_obs_5_15} and \ref{fig:VLBA_obs_22} using 
AIPS task JMFIT. We assumed 5\% uncertainties of these measurements. 
Table\,\ref{tab:fluxes} comprises the results along with the respective 
epochs to clarify which measurements are quasi-simultaneous and which are 
not. This is essential when trying to estimate the spectral indices because 
of the possible variability of the source's components which, if occurring, 
would complicate the interpretation of the results. However, regardless of 
the magnitude of variability of A2 between the epochs 2009.4 and 2011.0, it 
clearly has an inverted spectrum between 15.4\,GHz and 22\,GHz hence it most 
likely is the core. Combined with the unmeasurable 5\,GHz flux of A2, it 
appears that the flux density of A2 rises sharply in the whole range from 
5\,GHz up to 22\,GHz. Owing to this type of spectrum, A2 could be classified 
as a high-frequency peaker (HFP)
\citep[see the definition of that class in][]{Dallacasa2000}.

\begin{figure}[b]
\centering
\resizebox{\hsize-0.85cm}{!}{\includegraphics{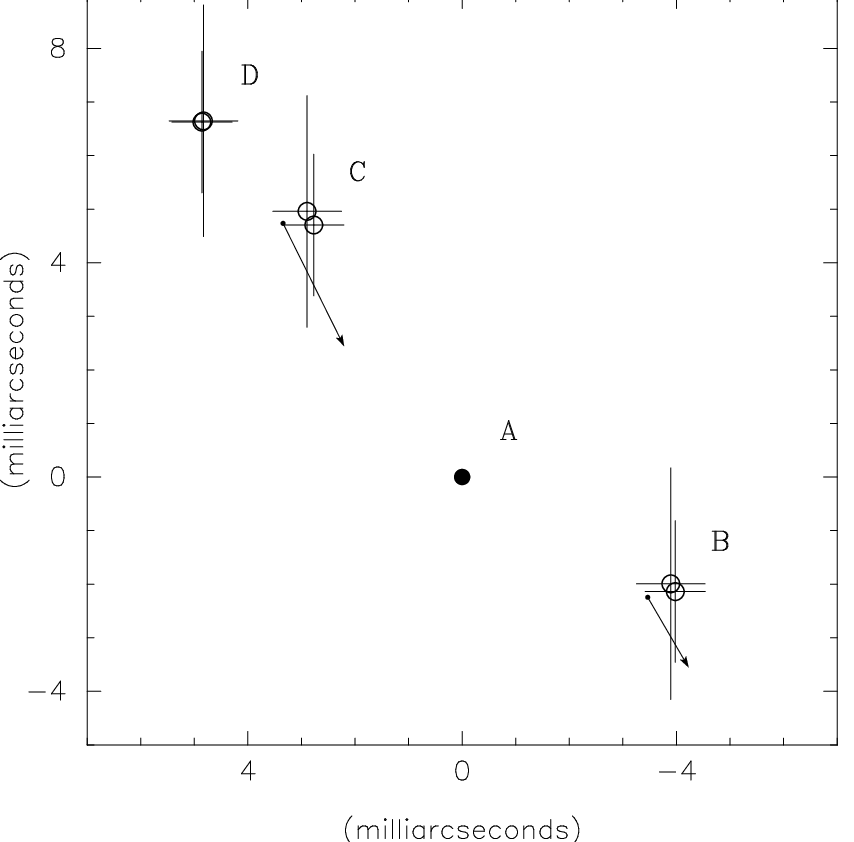}}
\caption{Schematic diagram showing the positions of the four components 
seen in the 5\,GHz images shown in Fig.\,\ref{fig:VLBA_obs_5_15}. Open circles
represent the positions of the centroids of B, C, and D with respect to A 
in the epochs 1997.6 and 2009.4. Error bars correspond to 20\% of the beam 
sizes. The arrows indicate the direction of the components' displacements. 
The length of each arrow is proportional to the magnitude of the respective 
displacement.}
\label{fig:model1}
\end{figure}

\section{Measurements of proper motions}
\label{sect:motions}

After we found the most likely candidate for the core, we attempted a 
crucial test of the CSO hypothesis: if the western part of 0932+075 was a 
CSO then like in most CSOs \citep[see][and references therein]{An2012} we 
would expect expansion of the mini-lobes B and D. By the same token, the 
component C would be expected to travel towards the north-east, i.e. in a 
way analogous to D since both are on the same side of the putative core. We 
measured the changes of the angular distances of B, C, and D with respect to 
A in the 5\,GHz images between the epochs 1997.6 and 2009.4 and we obtained 
the results which are graphically demonstrated in Fig.\,\ref{fig:model1} 
along with the corresponding uncertainties. The separation between A and D 
(8.2\,mas) remained nearly the same in both epochs, B moved away from A so 
that the angular distance between them increased from 4.4\,mas to 4.5\,mas, 
whereas C moved towards A -- the separation between these two decreased from 
5.7\,mas to 5.5\,mas. This combination of displacements cannot be reconciled 
with the hypothesis that the western part of 0932+075 is a CSO whose core 
belongs to the A-complex, because the non-core components of a CSO should 
either be stationary or, if not, then {\em all} of them should move away 
from the core.

Based on this surprising result, we rejected the hypothesis that the 
western part of 0932+075 is a CSO. Consequently, we claim that in the 5\,GHz 
VLBA images (Fig.\,\ref{fig:VLBA_obs_5_15}, upper panels) this object 
only mimics a CSO.

\begin{figure*}
\centering
\resizebox{\hsize/2-0.85cm}{!}{\includegraphics{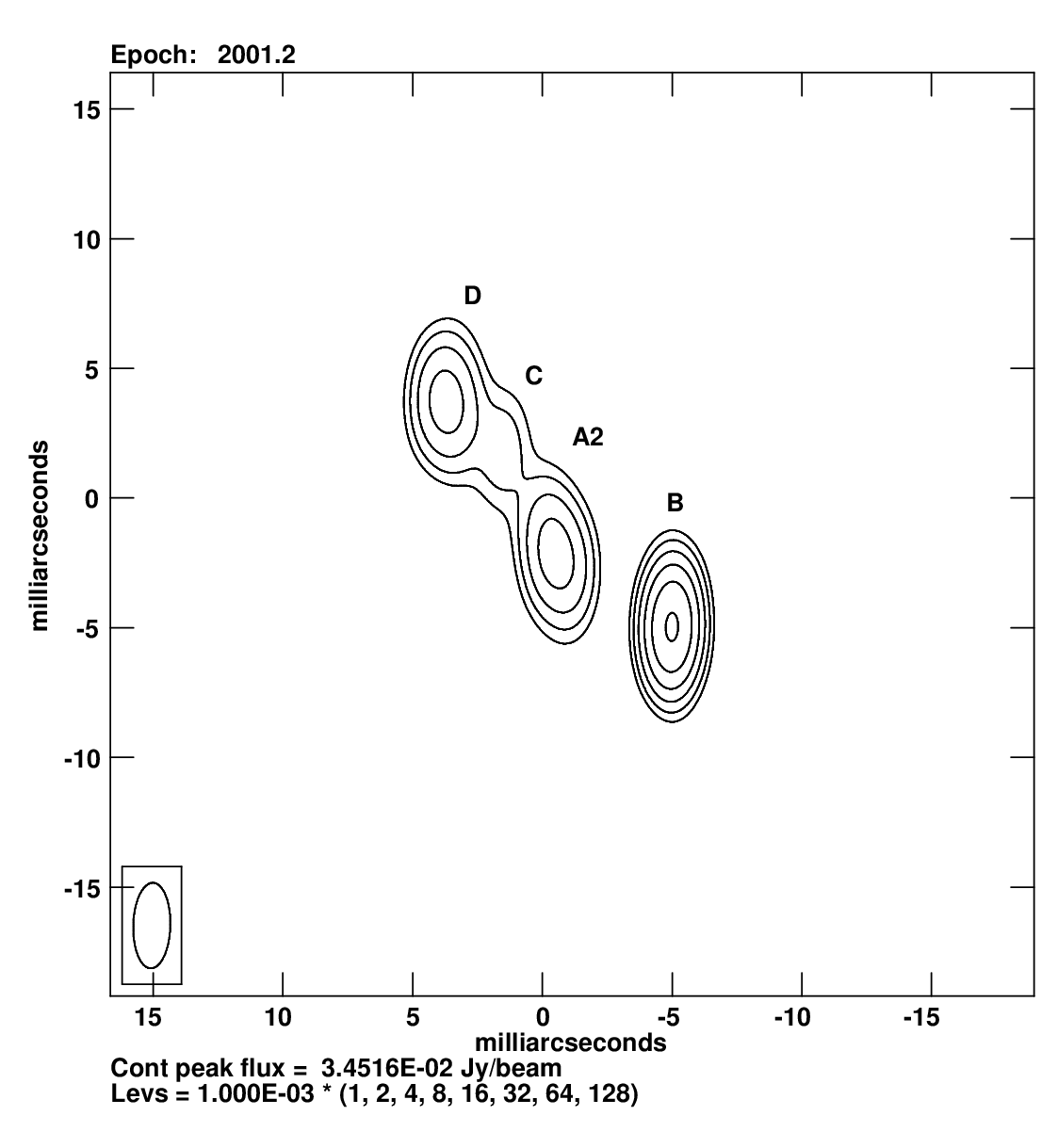}}
\resizebox{\hsize/2-0.85cm}{!}{\includegraphics{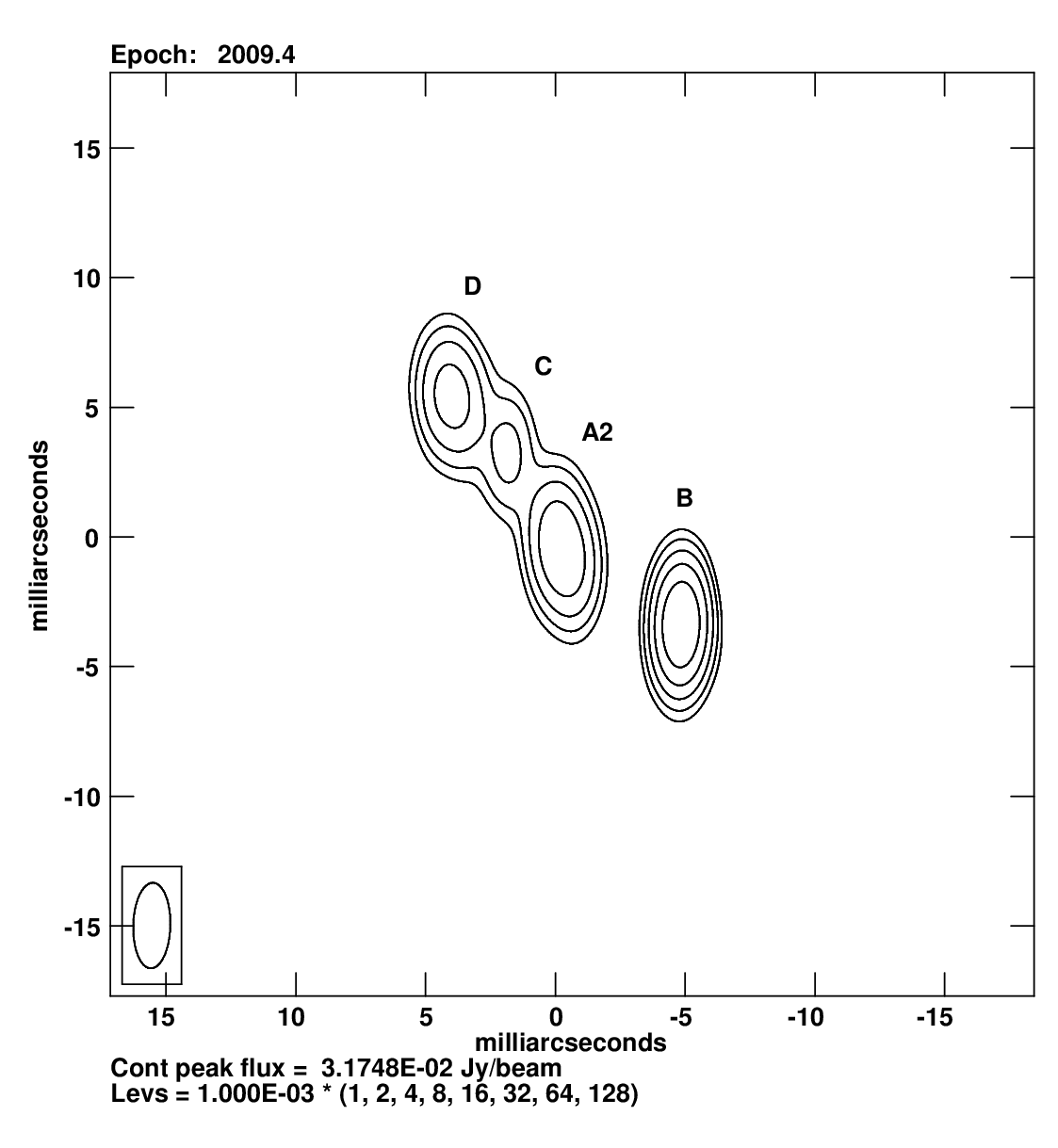}}
\caption{15.4\,GHz images convolved with the beam of the 5\,GHz observation
at the epoch 2009.4. {\em Left:} the image based on the observation in the 
epoch 2001.2, {\em Right:} the image based on the observation in the epoch 
2009.4.}
\label{fig:convolved}
\end{figure*}

\begin{figure*}
\centering
\resizebox{\hsize/2-0.85cm}{!}{\includegraphics{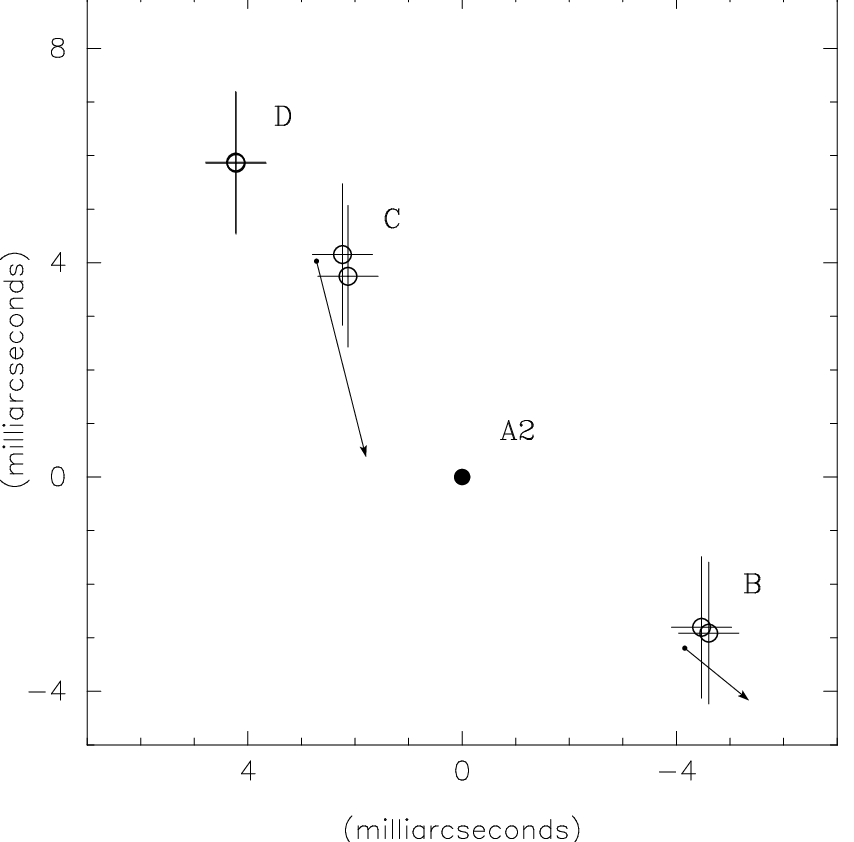}}
\resizebox{\hsize/2-0.85cm}{!}{\includegraphics{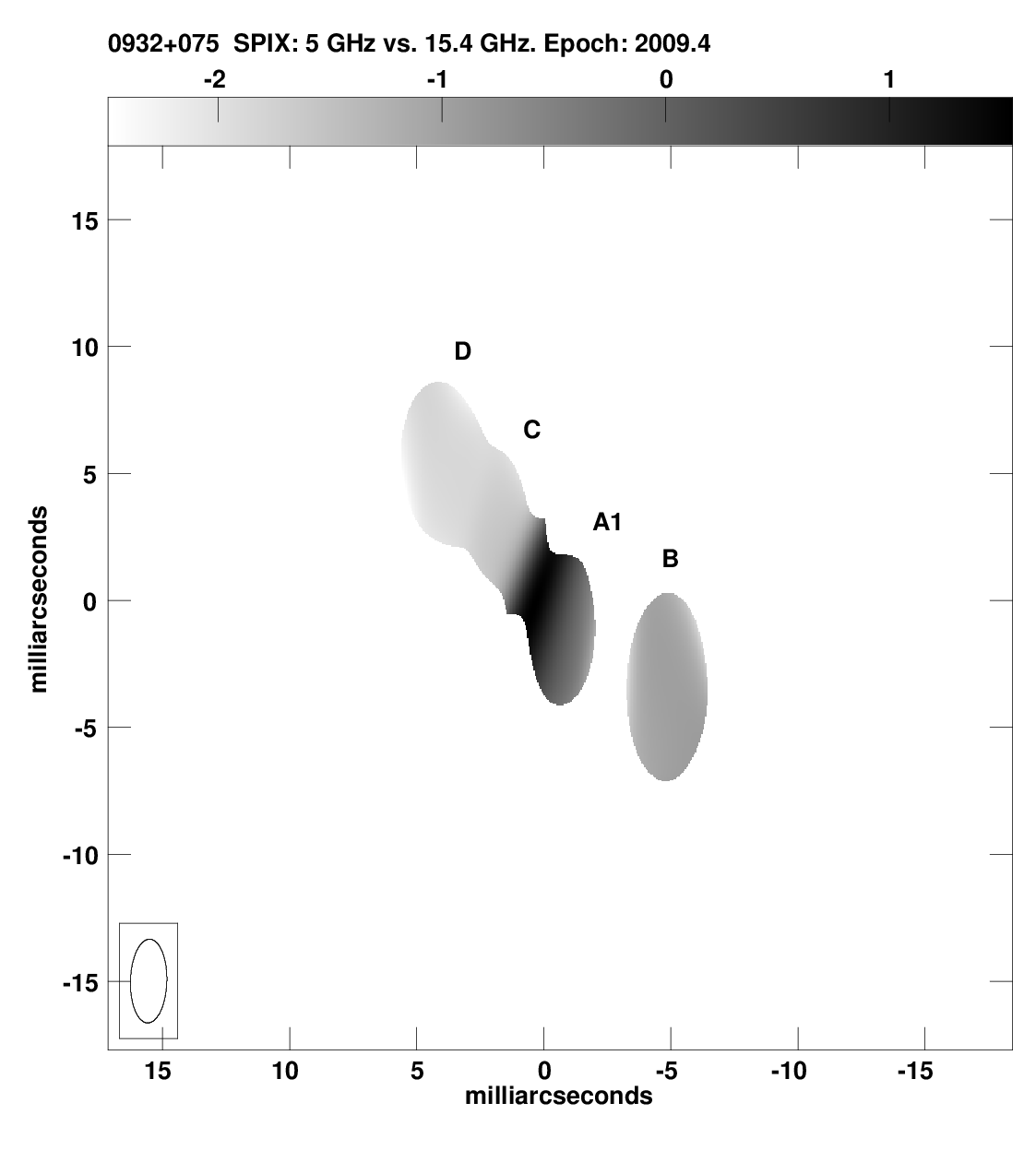}}
\caption{{\em Left:} Schematic diagram showing the positions of the four 
components seen in the 15.4\,GHz images convolved with a 5\,GHz beam shown 
in Fig.\,\ref{fig:convolved}. Open circles represent the positions of the 
centroids of B, C, and D with respect to A in the epochs 2001.2 and 2009.4. 
Error bars correspond to 20\% of the 5\,GHz beam sizes. The arrows indicate 
the direction of the components' displacements. The length of each arrow is 
proportional to the magnitude of the respective displacement. {\em Right:} 
5\,GHz vs. 15.4\,GHz spectral-index map for epoch 2009.4 prepared from the 
images shown in Fig.\,\ref{fig:VLBA_obs_5_15} (upper-right panel) and 
Fig.\,\ref{fig:convolved} (right panel).}
\label{fig:model2}
\end{figure*}

Given that the result we obtained is non-standard and taking the opportunity 
that there is another pair of high-resolution observations of 0932+075, i.e. 
the 15.4\,GHz VLBA observations of epochs 2001.2 and 2009.4 
(Fig.\,\ref{fig:VLBA_obs_5_15}, lower panels), we attempted a comparison 
between these two. However, to make the findings from the 15.4\,GHz data 
compatible with the 5\,GHz data described above, we convolved both 15.4\,GHz 
images with the beam of one of the 5\,GHz images, namely our image of epoch 
2009.4. The resulting smeared images are shown in Fig.\,\ref{fig:convolved}. 
We fitted Gaussian profiles to the features present in these maps and we 
measured the changes of the angular distances between component A (we note 
that it is now co-incident with A2) vs. components B, C, and D to find their 
displacements. The combined result is shown in Fig.\,\ref{fig:model2} (left 
panel). The separation between D and A2 remains constant over time which 
agrees with the 5\,GHz measurements. However, quantitatively it differs as 
it amounts to 7.2\,mas, i.e. $\sim$1\,mas less than in the case of the 
actual 5\,GHz maps because A2 in Fig.\,\ref{fig:model2} (left panel) has a 
different position than the component A in Fig.\,\ref{fig:model1} where it 
is co-incident with A1. For the same reason, the distances A2-B in the set 
of maps in Fig.\,\ref{fig:convolved} are greater by $\sim$1\,mas than those 
in the true 5\,GHz maps and amount to 5.3\,mas and 5.4\,mas in respective 
epochs. Furthermore, we have found that the A2-C distance has decreased 
between epochs 2001.2 and 2009.4 from 4.7\,mas to 4.3\,mas. This decrease 
confirms our earlier findings of the component C travelling in the opposite 
direction to the lobe expansion expected in the case of a CSO. Therefore, we 
again reached the conclusion that a CSO paradigm is not applicable to the 
case of 0932+075. The magnitude of that decrease is twice that measured at 
5\,GHz. This is somewhat uncomfortable, but tolerable given the 
uncertainties of all the above separations indicated by the error bars in 
the left panel of Fig.\,\ref{fig:model2}: their magnitudes are 20\% of the 
5\,GHz beam size.

At this stage, we took the opportunity that the 15.4\,GHz image of epoch 
2009.4 convolved with the 5\,GHz beam (Fig.\,\ref{fig:convolved}, right 
panel) is ideally suited to be combined with the 5\,GHz image of the same 
epoch (Fig.\,\ref{fig:VLBA_obs_5_15}, upper-right panel) to calculate the 
spectral indices between these two. The resulting spectral-index map is 
shown in Fig.\,\ref{fig:model2} (right panel). As can be clearly seen there,
the area featured by inverted spectrum, i.e. the location of the HFP, is not 
coincident with the centre of the silhouette of component A1. This is in 
full agreement with our findings reported in Sect.\,\ref{sect:epoch2}.

Finally, we compared the two original-beam 15.4\,GHz images of 0932+075. 
Here, the A2-B distance changed from 5.57$\pm$0.60\,mas to 
5.68$\pm$0.45\,mas with the uncertainties of 20\% of the 15.4\,GHz beam 
sizes. The difference between these figures remains in agreement with that 
derived from 5\,GHz maps and 15.4\,GHz maps convolved with a 5\,GHz beam. 
The A2-C distance has decreased from 4.29$\pm$0.73 to 4.00$\pm$0.55. This 
result is probably not as reliable as it might seem because of the 
noticeable departure from the Gaussian shape of C as seen in the epoch 
2001.2 map (Fig.\,\ref{fig:VLBA_obs_5_15}, lower-left panel) which causes 
difficulties with fitting a single Gaussian profile. Measuring the angular 
distance to D was problematic since it has a double structure at 15.4\,GHz 
(as well as at 22.2\,GHz). We failed to fit a Gaussian model to the 
component DW in the map from epoch 2001.2. Therefore, we only measured the 
separation A2-DE that changed from 7.31$\pm$0.71\,mas to 7.37$\pm$0.54\,mas. 
It follows that, while the D component as a whole remains stationary with 
respect to A, both at 5\,GHz and 15.4\,GHz, its subcomponents probably move 
apart. Second-epoch 22.2\,GHz observations where DE and DW are only 
sufficiently well resolved should shed more light on this question.

\section{Conclusions and future work}

Using the observational data presented here, we provided the two main pieces 
of new evidence characterising 0932+075. First, one of the components 
of the milliarcsecond-scale structure of this source (A2) has a clearly 
inverted spectrum for frequencies up to 22\,GHz. Owing to this property we 
label it a HFP. Second, it turns out that the western part of 0932+075 is
not a CSO as the 5\,GHz VLBA image of epoch 1997.6 (upper-left panel of
Fig.\,\ref{fig:VLBA_obs_5_15}) might suggest. Instead, we propose the 
following tentative interpretation of the milliarcsecond-scale morphology of 
the western part of 0932+075. What we see in the 15.4\,GHz and 22.2\,GHz VLBA 
images is a superposition of two groups of components. The first group 
consists of A2 and its closest neighbours A1 and A3, whereas the second 
group encompasses all the remaining components. This division was chosen
because the second group is nearly co-linear while the A-complex is an 
outlier. It is not clear what kind of relationship connects the HFP to the 
other components of 0932+075 and whether the co-linearity of the 
latter is significant, but there is a hint that it might be significant 
given that the directions of the displacements of B and C with respect to D are 
roughly the same as that of the line connecting these three components 
(see Fig.\,\ref{fig:model1}). It could be thus speculated that the co-linearity
is caused simply by the motion of B and C both travelling from the origin 
located at D. However, the weak point of this scenario is that D plays the 
role of a secondary core, which is doubtful because of its steep 
spectrum.

The main drawback of the observational material presented here is that Browne
et~al. did not use the phase referencing, and neither did we. The reason for
this is simple: 0932+075 itself is an acknowledged phase calibrator used for 
interferometric observations. Since we have now discovered quite complicated 
displacements in its milliarcsecond-scale structure, it appears that it 
would have been very advantageous if the VLBI observations of 0932+075 of 
all epochs had been phase referenced to the same phase calibrator so that 
the absolute positions of the components could have been known. 
Unfortunately, this is not the case and we had to rely on the measurements 
of the relative positions as presented in Sect.\,\ref{sect:motions}. In 
addition, the two-point proper motion measurements we have carried out are not 
very reliable hence the results presented here should be regarded as only 
preliminary. Consequently, a much more thorough study based on 
high-resolution monitoring of 0932+075 is necessary to carry out, as accurately 
as possible, measurements of the proper motions of its milliarcsecond-scale
components, based on their absolute positions obtained from phase referencing. 
The new observations should be multifrequency and quasi-simultaneous with 
regard to the set of frequencies at a particular epoch to find out what 
the exact spectral indices of each component are. Full polarisation 
information would also be helpful since only very few CSO jets have 
measurable linear polarisation, whereas in blazar jets fractional 
polarisation tends to increase with distance from the core.

\begin{acknowledgements}

The VLBA is operated by the National Radio Astronomy Observatory, a 
facility of the National Science Foundation operated under cooperative 
agreement by Associated Universities, Inc.\\

The VLBA observation of 0932+075 at 15.4\,GHz carried out on 22~March~2001 
was proposed by the CLASS team, specifically by Neal~J.~Jackson and Martin 
Norbury. It was devised to be an ultimate proof that 0932+075 was not a GL 
system. However, the image resulting from that observation has never been
published. The raw data belongs to the public domain yet we contacted the 
proposers and we have received written permission to publish the map resulting 
from that data in the present paper.\\

This research has made use of the NASA/IPAC Extragalactic Database (NED) 
which is operated by the Jet Propulsion Laboratory, California Institute of 
Technology, under contract with the National Aeronautics and Space 
Administration.

\end{acknowledgements}

\end{document}